\documentstyle[12pt]{article}
\topmargin -1.5cm \textheight 23cm \textwidth 160mm \vspace{1.8cm}
\begin{document}
\baselineskip 20pt
\begin{center}
\baselineskip=24pt {\Large \bf  Quantum wave equation of photon}

\vspace{1cm} {Xiang-Yao Wu$^{a}$
\footnote{E-mail:wuxy2066@163.com}, Xiao-Jing Liu$^{a}$, Yi-Heng
Wu$^{a}$\\ Qing-Cai Wang$^{a}$, Yan Wang$^{a}$ and Li-Xin
Chi$^{a}$}

\vspace{0.5cm}

\vskip 10pt \noindent{\footnotesize a. Institute of Physics, Jilin
Normal University, Siping 136000, China \\ \vskip 5pt}

\end{center}

\date{}

\renewcommand{\thesection}{Sec. \Roman{section}} \topmargin 10pt
\renewcommand{\thesubsection}{ \arabic{subsection}} \topmargin 10pt
{\vskip 5mm
\begin {minipage}{140mm}
\centerline {\bf Abstract}
\vskip 8pt
\par
\indent \hspace{0.3in}In this paper, we give the quantum wave
equations of single photon when it is in the vacuum and medium.
With these wave equations, we can study light interference and
diffraction with the approach of quantum theory, and also can
study the quantum property of photon when it is in a general
crystal and photonic crystal. Otherwise, it can be applied in
quantum optics and condensed matter field.
\end {minipage}

\vspace*{2cm} {\bf PACS number(s): 03.65.-w, 42.50.Ar, 42.70.Qs}


\newpage
\section * {1. Introduction}
\hspace{0.3in}It is known that one can describe single-photon
states using a photon-as-particle viewpoint, specifying the photon
wave function. The photon wave function and its equation of motion
are established from the Einstein energy-momentum-mass relation,
assuming a local energy density. According to modern quantum field
theory, photons, together with all other particles, are the
quantum excitations of a field. In the case of photons, these are
the excitations of the electromagnetic field. The lowest field
excitation of a given type corresponds to one photon and higher
field excitations involve more than one photon. This concept of a
photon enables one to use the photon wave function not only to
describe quantum states of an excitation of the free field but
also of the electromagnetic field interacting with a medium.

The photon wave function can not have all the properties of the
Schrodinger wave function of nonrelativistic wave mechanics.
Insistence on those properties that, owing to peculiarities of
photon dynamics, cannot be rendered, led some physicists to the
extreme opinion that the position probability density photon wave
function does not exist, and the photon wave function is
energy-density wave function in coordinate space, which has
developed over the past dozen years [1-4]. We know that the
quantized field theory of light developed by Dirac [5], and this
actually provides a derivation of the Maxwell equations, starting
from fundamental principles. The derivation parallels that of
Dirac for the electron and its quantum field [6, 7]. A key
difference between the electron and photon derivations has to do
with the famous localization problem for the photon [8]. Whereas
non-relativistic electrons can be in a position eigenstate, at
least in principle, a photon cannot. On the other hand, the energy
density of the electromagnetic field in free space can be
expressed as a local quantity, $E^{2} (x) + c^{2}B^{2} (x)$. Since
the photon wave function is the sum of real and imaginary parts,
i.e., $\vec{\psi}(\vec{r},t)=2^{-\frac{1}{2}}(\vec{E}(\vec{r},t)+i
\vec{H}(\vec{r},t))$. We call the mean-energy density wave
function or the Bialynicki-Birula-Sipe wave function and its
equation of motion is the photon wave equation. This section
introduces the photon wave function and its meaning. Sec. 2 gives
the potential energy of photon interaction with medium firstly and
give the energy of photon in medium. Sec. 3 gives the quantum wave
equation of single photon in the medium by extending the method as
Ref. [9]. Sec. 4 gives the application of quantum wave equation of
photon.

\section * {2. Relativistic Hamiltonian for a photon}

\hspace{0.3in} We know a relativistic Lagrangian for a particle in
external field is
\begin{equation}
L=-m_{0}c^{2}\sqrt{1-\frac{v^{2}}{c^{2}}}-V,
\end{equation}
where $m_{0}$ is the rest mass of the particle, $v$ is the
velocity of the particle, $c$ is the velocity of light and $V$ is
the potential energy of the particle in external field. The
canonical momentum $\vec{p}$ conjugate to the position coordinate
$\vec{x}$ is obtained by the definition,
\begin{equation}
p_{i}=\frac{\partial L}{\partial v_{i}}=\frac{m_{0}
v_{i}}{\sqrt{1-\frac{v^{2}}{c^{2}}}},
\end{equation}
and so
\begin{equation}
\vec{P}=\frac{m_{0}\vec{v}}{\sqrt{1-\frac{v^{2}}{c^{2}}}},
\end{equation}
the Hamiltonian takes on the form:
\begin{eqnarray}
H&=&E=\vec{P}\cdot\vec{v}-L=\frac{m_{0}c^{2}}{\sqrt{1-\frac{v^{2}}{c^{2}}}}+V\nonumber \\
&=&\sqrt{c^{2}p^{2}+m^{2}_{0}c^{4}}+V,
\end{eqnarray}
where $E$ is the total energy of the particle.

For a photon, the rest mass of $m_{0}=0$, and the (4) becomes
\begin{equation}
E=cp+V.
\end{equation}
In medium, the energy, momentum and velocity of photon are
\begin{equation}
E=h\nu, \hspace{0.15in} p=\frac{h}{\lambda},
\hspace{0.15in}v=\nu\lambda,
\end{equation}
where $\nu$ is photon frequency, and $\lambda$ is photon
wavelength. Substituting (6) into (5), we can obtain
\begin{equation}
hv=hc+V\lambda.
\end{equation}
From (7), we give the potential energy of photon in a medium
\begin{equation}
V=\frac{h(v-c)}{\lambda}=\frac{h(\frac{c}{n}-c)}{\lambda}=\frac{hc}{\lambda}(\frac{1}{n}-1).
\end{equation}
Substituting (8) into (5), we can give the energy of photon in
medium
\begin{equation}
E=cp+V=pv.
\end{equation}

\section * {3. Quantum wave equation of photon in medium}

\hspace{0.3in} With the photon energy equation (5), we can study
the quantum wave equation of photon in medium. The equation (5)
can be written as
\begin{equation}
E=c\sqrt{\vec{p}\cdot\vec{p}}+V.
\end{equation}
Define a multicomponent wave function $\vec{\psi}(\vec{p},E)$
obeying the normalization condition
\begin{equation}
(2\pi\hbar)^{-3}\int d^{3}p \vec{\psi}^{*}(\vec{p},E)\cdot
\vec{\psi}(\vec{p},E)=1.
\end{equation}
Since a photon spin $s=1$, its wave function has three components,
i.e., $\vec{\psi}(\vec{p},E)=(\psi_{x}(\vec{p},E),
\psi_{y}(\vec{p},E), \psi_{z}(\vec{p},E))$, we multiply (10) by
the wave function $\vec{\psi}(\vec{p},E)$ to give
\begin{equation}
E
\vec{\psi}(\vec{p},E)=(c\sqrt{\vec{p}\cdot\vec{p}}+V)\psi(\vec{p},E).
\end{equation}
In order to represent the square-root operator $\sqrt{\vec{p}
\cdot \vec{p}}$, we define a vector operator
$\hat{A}=i\vec{p}\times$, where $\times$ is the cross product
operator[9], we have
\begin{equation}
\hat{A}\hat{A}\vec{\psi}=-\vec{p}\times(\vec{p}\times\vec{\psi})=
(\vec{p}\cdot\vec{p})\vec{\psi}-\vec{p}(\vec{p}\cdot\vec{\psi}).
\end{equation}
Any vector field can be written as the sum of two linearly
independent part, $\vec{\psi}=\vec{\psi}_{T}+\vec{\psi}_{L}$,
where the transverse part $\vec{\psi}_{T}$ and longitudinal part
$\vec{\psi}_{L}$ obey
\begin{equation}
\vec{p}\cdot\vec{\psi}_{T}=0,
\hspace{0.15in}\vec{p}\times\vec{\psi}_{L}=0,
\end{equation}
and (13) becomes
\begin{equation}
\hat{A}\hat{A}\vec{\psi}=
\vec{p}\cdot\vec{p}(\vec{\psi}_{T}+\vec{\psi}_{L})-
\vec{p}(\vec{p}\cdot(\vec{\psi}_{T}+\vec{\psi}_{L})).
\end{equation}
Identifying the transverse part $\vec{\psi_{L}}$ as the relevant
field for the photon. For the photon field $\vec{\psi_{L}}$, (15)
becomes
\begin{equation}
\hat{A}\hat{A}\vec{\psi}_{T}=\vec{p}\cdot\vec{p}\vec{\psi}_{T},
\end{equation}
and then
\begin{equation}
\sqrt{\hat{A}\hat{A}}\vec{\psi}_{T}=\hat{A}\vec{\psi}_{T}=\sqrt{\vec{p}\cdot\vec{p}}\vec{\psi_{T}},
\end{equation}
we multiply (17) by the light velocity $c$ and substitute (12)
into (17) to give
\begin{equation}
c\hat{A}\vec{\psi}_{T}=c\sqrt{\vec{p}\cdot\vec{p}}\vec{\psi}_{T}=(E-V)\vec{\psi}_{T},
\end{equation}
or
\begin{equation}
(E-V)\vec{\psi}_{T}(\vec{p},
E)=ci\vec{p}\times\vec{\psi}_{T}(\vec{p}, E),
\end{equation}
From (19), we know $\vec{\psi}_{T}(\vec{p},E)$ must be a
complex-valued vector, we can make Fourier transform for wave
function $\vec{\psi}(\vec{p},E)$ from momentum space to coordinate
space, and from energy to time, accounting for the constraint
between energy and momentum ($E=c|\vec{p}|$) by including a delta
function. The energy $E$ to be considered as an independent
variable, and gives [9]
\begin{equation}
\vec{\psi}(\vec{r},t)=(2\pi\hbar)^{-4}\int\int dE d^{3}p\delta
(E-v|\vec{p}|)exp(-iEt/\hbar+i\vec{p}\cdot\vec{r}/\hbar)f(E)\vec{\psi}(\vec{p},E).
\end{equation}
The weight function $f(E)$ has been included to allow different
forms of normalization of the coordinate -space function
$\vec{\psi}_{T}(\vec{r}, t)$. For the photon, we adopt the choice
in Ref. [9], which gives for the coordinate-space normalization,
\begin{equation}
\int d^{3}r\vec{\psi}^{*}(\vec{r},t)\cdot\vec{\psi}(\vec{r},t)=
(2\pi\hbar)^{-3}\int
d^{3}pE(p)\vec{\psi}^{*}(\vec{p},E)\cdot\vec{\psi}(\vec{p},E)=\langle
E\rangle,
\end{equation}
where the $\langle E\rangle$ denotes the expectation value of the
photon's energy. For the photon wave function
$\vec{\psi}_{T}(\vec{r}, t)$, there is the same transform as
$\vec{\psi}(\vec{r}, t)$
\begin{equation}
\vec{\psi}_{T}(\vec{r},t)=(2\pi\hbar)^{-4}\int\int dE d^{3}p
f(E)\delta
(E-v|\vec{p}|)exp(-iEt/\hbar+i\vec{p}\cdot\vec{r}/\hbar)\vec{\psi}_{T}(\vec{p},E).
\end{equation}
From (22), there are the reversal Fourier transforms as follows:
\begin{eqnarray}
&&E\cdot
f(E)\delta(E-v|\vec{p}|)\vec{\psi_{T}}(\vec{p},E)\nonumber\\&=&(2\pi\hbar)^{-4}\int\int
dtd^{3}r E \cdot exp(iEt/\hbar-i\vec{p}\cdot\vec{r}/\hbar)\vec{\psi}_{T}(\vec{r},t)\nonumber\\
\nonumber\\&=&(2\pi\hbar)^{-4}\int\int
dtd^{3}r\frac{h}{i}\frac{\partial}{\partial
t}\{exp[iEt/\hbar-i\vec{p}\cdot\vec{r}/\hbar]\}\vec{\psi}_{T}(\vec{r},t)\nonumber\\
&=&(2\pi\hbar)^{-4}\frac{h}{i}\{\int\int
dtd^{3}r(\frac{\partial}{\partial
t}[exp(iEt/\hbar-i\vec{p}\cdot\vec{r}/\hbar)\cdot\vec{\psi}_{T}(\vec{r},t)]\nonumber\\&&-
exp(iEt/\hbar-i\vec{p}\cdot\vec{r}/\hbar)\frac{\partial}{\partial
t}\vec{\psi}_{T}(\vec{r},t))\}\nonumber\\
&=&(2\pi\hbar)^{-4}\frac{h}{i}\{\int\int
d^{3}rd[exp(iEt/\hbar-i\vec{p}\cdot\vec{r}/\hbar)\cdot\vec{\psi}_{T}(\vec{r},t)]\nonumber\\&&-
\int\int dtd^{3}r exp(i E
t/\hbar-i\vec{p}\cdot\vec{r}/\hbar)\frac{\partial}{\partial
t}\vec{\psi}_{T}(\vec{r},t)\}\nonumber\\&=&(2\pi\hbar)^{-4}i\hbar\int\int
dtd^{3}rexp(iEt/\hbar-i\vec{p}\cdot\vec{r}/\hbar)\frac{\partial}{\partial
t}\vec{\psi}_{T}(\vec{r},t),
\end{eqnarray}
\begin{eqnarray}
&&\delta(E-v|\vec{p}|)f(E)\vec{p}\times\vec{\psi}_{T}(\vec{p},E)\nonumber\\
&=&(2\pi\hbar)^{-4}\int\int
dtd^{3}r exp(i E t/\hbar-i\vec{p}\cdot\vec{r}/\hbar)\vec{p}\times\vec{\psi}_{T}(\vec{r},t)\nonumber\\
&=&(2\pi\hbar)^{-4}\int\int
dtd^{3}r\bigtriangledown[exp(iEt/\hbar-i\vec{p}\cdot\vec{r}/\hbar)](-\frac{h}{i})\times\vec{\psi}_{T}(\vec{r},t)\nonumber\\
&=&-(2\pi\hbar)^{-4}\frac{h}{i}\int\int
dtd^{3}r(\bigtriangledown[exp(iEt/\hbar-i\vec{p}\cdot\vec{r}/\hbar)\times\vec{\psi}_{T}(\vec{r},t)]\nonumber\\&&-
exp(iEt/\hbar-i\vec{p}\cdot\vec{r}/\hbar)\bigtriangledown\times\vec{\psi}_{T}(\vec{r},t))\nonumber\\
&=&-(2\pi\hbar)^{-4}i\hbar\int\int
dtd^{3}rexp(iEt/\hbar-i\vec{p}\cdot\vec{r}/\hbar)\bigtriangledown\times\vec{\psi}_{T}(\vec{r},t),
\end{eqnarray}
and
\begin{eqnarray}
&&\delta(E-v|\vec{p}|)f(E)V\vec{\psi}_{T}(\vec{p},E)\nonumber\\&=&(2\pi\hbar)^{-4}\int\int
dtd^{3}rexp(iEt/\hbar-i\vec{p}\cdot\vec{r}/\hbar)V\vec{\psi}_{T}(\vec{r},t)
\nonumber\\
&=&(2\pi\hbar)^{-4}i\hbar\int\int dt d^{3} r exp(i E t /
\hbar-i\vec{p}\cdot\vec{r}/\hbar)\frac{V}{i\hbar}\vec{\psi}_{T}(\vec{r},t),
\end{eqnarray}
substituting (23), (24) and (25) into (19), we have
\begin{equation}
i\hbar\frac{\partial}{\partial
t}\vec{\psi}_{T}(\vec{r},t)=c\hbar\nabla\times\vec{\psi}_{T}(\vec{r},t)+V\vec{\psi}_{T}(\vec{r},t).
\end{equation}
Equation (26) is quantum wave equation of photon in medium.

For the free photon ($V=0$), we have
\begin{equation}
i\frac{\partial}{\partial
t}\vec{\psi}_{T}(\vec{r},t)=c\nabla\times\vec{\psi}_{T}(\vec{r},t).
\end{equation}
Equation (27) is quantum wave equation of free photon, and it is
the same as Ref. [9]. Multiplying (27) by $\hbar$, we obtain
\begin{equation}
i\hbar\frac{\partial}{\partial
t}\vec{\psi}_{T}(\vec{r},t)=c\hbar\nabla\times\vec{\psi}_{T}(\vec{r},t)
=c(\hat{\vec{p}}\cdot\vec{s})\vec{\psi}_{T}(\vec{r},t),
\end{equation}
with
\begin{eqnarray}
s_{x}= \left ( \begin{array}{ll}
0 \hspace{0.2in}   0 \hspace{0.2in}  0 \\
0 \hspace{0.2in}   0 \hspace{0.1in}  -i\\
0 \hspace{0.2in}   i \hspace{0.2in}  0
 \end{array}
  \right  ),
s_{y}= \left ( \begin{array}{ll}
0 \hspace{0.2in}    0 \hspace{0.2in}  i \\
0 \hspace{0.2in}    0 \hspace{0.2in}  0\\
-i \hspace{0.1in}   0 \hspace{0.2in}  0
 \end{array}
  \right  ),
s_{z}= \left ( \begin{array}{ll}
0 \hspace{0.1in}  -i \hspace{0.1in}  0\\
i \hspace{0.2in}    0 \hspace{0.2in}  0\\
0 \hspace{0.2in}   0 \hspace{0.2in}  0
 \end{array}
  \right  ),
\end{eqnarray}
we find
\begin{equation}
s^{2}=s_{x}^{2}+s_{y}^{2}+s_{z}^{2}=2\left ( \begin{array}{ll}
1 \hspace{0.2in}  0 \hspace{0.2in}  0\\
0 \hspace{0.2in}  1 \hspace{0.2in}  0\\
0 \hspace{0.2in}  0 \hspace{0.2in}  1
 \end{array}
  \right  )=2I=s(s+1)I.
\end{equation}
From (30), we have $s=1$. So, (27) and (28) are the particle wave
equation, corresponding to spin $s=1$ and rest mass $m_{0}=0$,
i.e., (27) and (28) are the quantum wave equation of photon.

\section * {4. The application of the photon quantum wave equation}

\hspace{0.2in} In the following, we give some examples about the
application of the photon quantum wave equation, which include to
study the quantum property of light in general medium and in
photonic crystal. We think the equation can be applied in many
fields.

If we suppose the medium there are $N$ molecules per unit volume
with $z$ electrons per molecule, and that, instead of a single
binding frequency for all, there are $f_{s}$ electrons per
molecule with binding frequency $\omega _{s}$ and damping constant
$\gamma_{s}$, then the medium index $n$ is given by
\begin{equation}
n^{2}=1+\frac{Ne^{2}}{\varepsilon_{0}m}\sum\limits_{s}\frac{f_{s}}{\omega_{s}^{2}-\omega^{2}-i\omega\gamma_{s}},
\end{equation}
where $\omega$ is the frequency of incident photon, $m$ is the
electron mass. Substituting (31) into (8), we can obtain the
potential energy of photon in medium
\begin{equation}
V=\frac{hc}{\lambda}(\frac{1}{n}-1)=
\frac{hc}{\lambda}[(1+\frac{Ne^{2}}{\varepsilon_{0}m}\sum\limits_{s}
\frac{f_{s}}{\omega_{s}^{2}-\omega^{2}-i\gamma_{s}\omega})^{-\frac{1}{2}}-1].
\end{equation}
Substituting (32) into (26), we can obtain the quantum wave
equation of photon in medium
\begin{equation}
i\hbar\frac{\partial}{\partial
t}\vec{\psi}_{T}(\vec{r},t)=c\hbar\nabla\times\vec{\psi}_{T}(\vec{r},t)+
\frac{hc}{\lambda}[(1+\frac{Ne^{2}}{\varepsilon_{0}m}\sum\limits_{s}
\frac{f_{s}}{\omega_{s}^{2}-\omega^{2}-i\gamma_{s}\omega})^{-\frac{1}{2}}-1]\vec{\psi}_{T}(\vec{r},t).
\end{equation}
The equation (33) can be solved by the method of separation of
variable. By writing
\begin{equation}
\vec{\psi}(\vec{r},t)=\vec{\psi}(\vec{r})g(t),
\end{equation}
substituting (34) into (33), we have
\begin{equation}
g(t)=e^{-\frac{i}{\hbar}Et},
\end{equation}
and
\begin{equation}
c\hbar\nabla\times\vec{\psi}_{T}(\vec{r})=\hbar\omega [2-
(1+\frac{Ne^{2}}{\varepsilon_{0}m}\sum\limits_{s}
\frac{f_{s}}{\omega_{s}^{2}-\omega^{2}-i\gamma_{s}\omega})^{-\frac{1}{2}}]\vec{\psi}_{T}(\vec{r}),
\end{equation}
simplifying (36), we have
\begin{equation}
\nabla\times\vec{\psi}_{T}(\vec{r})=k[2-(1+\frac{Ne^{2}}{\varepsilon_{0}m}\sum\limits_{s}
\frac{f_{s}}{\omega_{s}^{2}-\omega^{2}-i\gamma_{s}\omega})^{-\frac{1}{2}}]\vec{\psi}_{T}(\vec{r}).
\end{equation}
From (37), we can study the quantum property of light in medium.

When the refractive index of medium is periodical in space, the
medium is photonic crystal, and its refractive index can be
written as:
\begin{equation}
n(\vec{r})=n(\vec{r}+m\tau)\hspace{0.5in} m=0, 1,
2\cdot\cdot\cdot,
\end{equation}
where $\tau$ is period. Substituting (38) and (8) into (26), we
have
\begin{equation}
i\hbar \frac{\partial}{\partial t}\vec{\psi}_{T}(\vec{r},
t)=c\hbar \nabla\times\vec{\psi}_{T}(\vec{r},
t)+\frac{hc}{\lambda}(\frac{1}{n(\vec{r})}-1)\vec{\psi}_{T}(\vec{r},
t),
\end{equation}
the (39) is quantum wave equation of photon in photonic crystal.
From the equation, we can study the quantum property of light in
photonic crystal.

\section * {5. Conclusion}

\hspace{0.2in} In conclusion, we firstly give the potential energy
of photon interaction with medium, and also give the photon
quantum wave equation in the vacuum and medium. With these quantum
wave equations, we can study light interference and diffraction,
and can study the quantum property of light in general medium and
photonic crystal.

\newpage
\end{document}